# Learning to Detect Fake Face Images in the Wild

Chih-Chung Hsu[#], Chia-Yen Lee[*], Yi-Xiu Zhuang[#]

[#]*Department of Management Information Systems, National Pingtung University of Science and Technology*
*1, Shuefu Road, Neipu, Pingtung 91201, Taiwan*
[1]`cchsu@mail.npust.edu.tw`
[3]`shaun85528@gmail.com`

[*]*Department of Electrical Engineering, National United University*
*2, Lienda, Miaoli 36063, Taiwan*
[2]`leecyya@gmail.com`

*Abstract*— **Although Generative Adversarial Network (GAN) can be used to generate the realistic image, improper use of these technologies brings hidden concerns. For example, GAN can be used to generate a tampered video for specific people and inappropriate events, creating images that are detrimental to a particular person, and may even affect that personal safety. In this paper, we will develop a deep forgery discriminator (DeepFD) to efficiently and effectively detect the computer-generated images. Directly learning a binary classifier is relatively tricky since it is hard to find the common discriminative features for judging the fake images generated from different GANs. To address this shortcoming, we adopt contrastive loss in seeking the typical features of the synthesized images generated by different GANs and follow by concatenating a classifier to detect such computer-generated images. Experimental results demonstrate that the proposed DeepFD successfully detected 94.7% fake images generated by several state-of-the-art GANs.**

*Keywords*— forgery detection, GAN, Contrastive loss, deep learning, fully convolutional network

## I. INTRODUCTION

With the rapid growth of deep learning techniques for image generation, the security issues because of malicious uses of such image generation/synthesis became more important. For example, progressive growing of GANs (PGGAN) proposed by nVidia [4] have demonstrated that the realistic and high-resolution face images can be easily synthesized. It can be used to create a fake personal account on Facebook to cheat something or someone [1]. It will cause very serious problems on the society, political, and commercial activities. Therefore, an effective and efficient images forgery detection technique is desired.

To address the issue of image forgery detection, there are two different categories in the traditional approach: 1) extrinsic feature and 2) intrinsic feature. First forgery detection will embed external unique signals into the original images (e.g. digital watermarking). Then, the received image can be verified whether it is a forgery or not by comparing the extracted watermark and the original watermark. The second strategy will tend to discover the intrinsic and invariant features from the original images. The forgery image should be able to detected by checking the statistical property of the extracted intrinsic feature from the received image because any tampered operation will make the intrinsic feature changed. The first strategy needs the original externally signal to check whether it is a forgery or not. In general, it is relatively hard to have originally external signal (i.e. watermark) for any received image. In contrast, the second strategy only seeks the intrinsic feature of the received image, finding the unusual statistical property of it, to detect whether it is a forgery or not [3].

There are several approaches to find the intrinsic features of images to determine the tampered images [2][3] The forgery detection technique in [2] finds sensor pattern noise as the intrinsic feature. Double compression cues are used in [13] as the intrinsic feature for JPEG formatted image. However, traditional forgery detection techniques are hard to detect the generated images by GANs since their image content are made by deep neural network directly. Therefore, it does not exist any unusually statistical property in the intrinsic features of the received images, leading to traditional forgery detection approach fails to detect the generated images. To solve this shortcoming, we propose a deep neural network called deep forgery discriminator (DeepFD) based strategy to effectively and efficiently detect the generated / fake images synthesized by GANs or other advanced networks.

There is an easy to train a deep neural network classifier to distinguish fake images and real images by collecting a large training set containing fake and real images. However, the trained classifier might be ineffectively to detect the fake images synthesized by a new GAN or other advanced generators because it did not learn the discriminative features from newly generated images. In general, it is hard to collect training images generated by all possible GANs or image synthesizers. Moreover, there are many new GANs proposed every year. Such strategy needs to re-train the classifier to keep its performance when there is a new GAN proposed. To ensure the performance of the proposed DeepFD, we tend to learn the jointly discriminative features from collected training images across different GANs by introducing the contrastive loss into the network learning framework [12]. Contrastive loss learns such joint features from heterogeneous training images by introducing the pairwise information so that the DeepFD should be able to effectively distinguish any fake image generated by any GAN. Besides, the proposed DeepFD can further localize unrealistic details of the fake image based on a fully convolutional architecture. Our contributions are summarized below:

- We propose a novel deep neural network based discriminator based on contrastive loss, upon which we can distinguish the fake images generated by any GANs. To the best of our knowledge, the proposed method is the first research effectively addressing generated image detection.
- The proposed DeepFD can be used to localize unrealistic details of the fake image and we can follow such regions to further improve the performance of the proposed DeepFD.

The rest of this paper is organized as follows. Sec. II presents the proposed DeepFD for forgery image detection. In Sec. III, experimental results are demonstrated. Finally, conclusions are drawn in Sec. IV.

## II. THE PROPOSED DEEP FORGERY DISCRIMINATOR

Figure 1 shows the flowchart of the proposed DeepFD. In the proposed method, we have two learning phases. First, we collect a lot of fake images synthesized by several GANs called example-generative model and real images to learn the jointly discriminative features $D_1$ based on the proposed contrastive loss. Afterward, a discriminator (classifier) $D_2$ will be concatenated to the $D_1$ to further distinguish fake images. In the test phase, it is easy to distinguish the test image whether it is fake or real by $D_1$ and $D_2$ directly. The details of network architectures in $D_1$ and $D_2$ are respectively described in Table I. Note, the classifier $D_2$ is directly concatenated to the 4$^{th}$ layer of the jointly discriminative feature learning network $D_1$.

Let the training set collected from $M$ GANs be $\mathbf{X}_{fake} = [\mathbf{x}_{i=1}^{k=1}, \mathbf{x}_{i=2}^{k=1}, \dots, \mathbf{x}_{i=N_1}^{k=1}, \dots, \mathbf{x}_{i=N_M}^{k=M}]$, where each GAN will generate $N_k$ training images. Let the real training set be $\mathbf{X}_{real} = [\mathbf{x}_{i=1}, \mathbf{x}_{i=2}, \dots, \mathbf{x}_{i=N_r}]$ containing $N_r$ training images. Total number of training images containing real and fake images will be $N_T = N_r + N_f = N_r + \sum_{k=1}^{M} N_k$. The label Y = $[y_1, y_2, \dots, y_{N_T}]$ indicates the image is fake (y=0) or real (y=1). To integrate the pairwise information into learning architecture, there are $C(N_T, 2)$ pairs in the pairwise information P = $[p_{i=0,j=0}, p_{i=0,j=1}, \dots, p_{i=0,j=N_r}, \dots, p_{i=N_f,j=N_r}]$. In general, it is unnecessary to generate all possible pairs to learn the jointly discriminative features of the fake images. In our experiments, it is enough that we adopt 1 million pairs to learn the joint feature.

### A. Jointly Discriminative Feature Learning

Let the feature representation $R_i = D_1(\mathbf{x}_i)$. With a paired input images, the purpose of jointly discriminative feature learning is to minimize the similarity function defined as follows:

$$E_W(\mathbf{x}_1, \mathbf{x}_2) = \|D_1(\mathbf{x}_1) - D_1(\mathbf{x}_2)\|,$$

where we use the exact same network $D_1$ to extract the features of the paired input images. Directly minimizing $E_W(\mathbf{x}_1, \mathbf{x}_2)$ may lead the feature representation $D_1(\mathbf{x}_i)$ to a constant mapping [12]. This situation will make the feature representation ineffective and useless. In contrast, we introduce the contrastive loss:

$$L(W, (P, \mathbf{x}_1, \mathbf{x}_2))$$

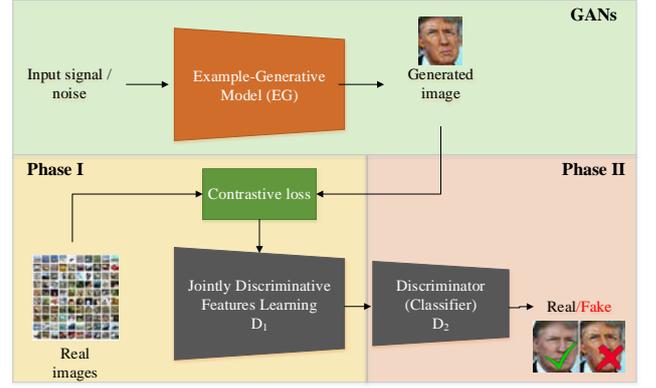

Fig. 1 The flowchart of the proposed deep forgery discriminator (DeepFD). Example-generative model (one or more than one models) will continuously synthesize training samples for the proposed DeepFD training.

TABLE I
NETWORK STRUCTURES IN JOINTLY DISCRIMINATIVE FEATURE LEARNING ($D_1$) AND CLASSIFIER TRAINING ($D_2$)

| Layers | $D_1$ | $D_2$ |
|---|---|---|
| 1 | Conv.layer, kernel=7*7, stride=4, channel=96 | Conv. layer, kernel=3*3, channel = 2 |
| 2 | Residual block *2, channel=96 | Global average pooling |
| 3 | Residual block *2, channel=128 | Fully connected layer, neurons=2 Softmax layer |
| 4 | Residual block *2, channel=256 | |
| 5 | Fully connected layer, neurons=128 Softmax layer | |

$$= \tfrac{1}{2}\big(p_{ij}(E_W)^2 + (1 - p_{ij})(\max(0, m - E_W)^2\big),$$

where $E_W = \|D_1(\mathbf{x}_1) - D_1(\mathbf{x}_2)\|$ and $m$ is the predefined marginal value. In this manner, it is possible to learn the common characteristic of the fake images generated by different GANs. With the contrastive loss, the feature representation $D_1(\mathbf{x}_i)$ will tend to become similar to $D_1(\mathbf{x}_j)$ if $p_{ij} = 1$ (i.e., fake-fake or real-real pair). By iteratively train the network $D_1$ based on the contrastive loss, the jointly discriminative feature should be able to be well learned.

### B. Classifier Training

Once the jointly discriminative feature representation has learned, there are several ways to classify the received images such as SVM, random forest classifier, or Bayer classifier. In this work, we directly concatenate a convolutional layer and a fully connected layer to the network $D_1$ (See details in Table I). In this way, the proposed DeepFD will be an end-to-end architecture.

The loss function of the classifier can be defined as a cross-entropy loss:

TABLE II
THE PERFORMANCE COMPARISON BETWEEN THE PROPOSED DEEPFD AND OTHER METHODS

| Method | Exclusive of LSGAN | | Exclusive of DCGAN | | Exclusive of WGAN | | Exclusive of WGAN-GP | | Exclusive of PGGAN | |
|---|---|---|---|---|---|---|---|---|---|---|
| | Precision | Recall | Precision | Recall | Precision | Recall | Precision | Recall | Precision | Recall |
| Traditional method [3] | 0.205 | 0.580 | 0.253 | 0.774 | 0.235 | 0.673 | 0.242 | 0.604 | 0.222 | 0.862 |
| CNN+RFC | 0.547 | 0.566 | 0.563 | 0.585 | 0.546 | 0.603 | 0.580 | 0.536 | 0.520 | 0.531 |
| CNN+SVM | 0.614 | 0.630 | 0.561 | 0.593 | 0.570 | 0.589 | 0.602 | 0.563 | 0.547 | 0.570 |
| CNN+LC | 0.599 | 0.567 | 0.547 | 0.521 | 0.563 | 0.539 | 0.582 | 0.583 | 0.500 | 0.474 |
| DeepFD w/o contrastive loss | 0.836 | 0.801 | 0.760 | 0.721 | 0.724 | 0.722 | 0.706 | 0.687 | 0.759 | 0.766 |
| The proposed DeepFD | **0.947** | **0.922** | **0.871** | **0.844** | **0.838** | **0.847** | **0.818** | **0.835** | **0.926** | **0.918** |

$$L_C(\mathbf{x}_i, y_i) = -\sum_i^{N_T} \big(D_2(D_1(\mathbf{x}_i))\log y_i\big).$$

The classifier can be easily trained by back-propagation [10]. Note, the parameters in all layers in $D_1$ will be updated as well because the jointly discriminative feature learned by $D_1$ might be further improved. In this way, we believe that the fake image can be effectively detected.

### III. EXPERIMENTAL RESULTS

#### A. Experimental Settings

To generate the training samples, the CelebA dataset will be used in this experiments. The images in CelebA cover large pose variations and background clutter including 10,177 number of identities and 202,599 aligned face images.

In experiments, we collected five state-of-the-art GANs to generate the fake images pool:
1) DCGAN (Deep convolutional GAN) [6]
2) WGAP (Wasserstein GAN) [7]
3) WGAN-GP (WGAN with Gradient Penalty) [8]
4) LSGAN (Least Squares GAN) [9]
5) PGGAN [4]

Each GAN will be used to generate 200,000 fake images sized of $64 \times 64$ into the fake image pool. PGGAN will produce a high-resolution result sized of $64 \times 64$. We downsample the fake images generated by PGGAN into resolution $64 \times 64$ to have a fair comparison. Since the resolution of fake image of PGGAN is downscaled into $64 \times 64$, the recognition task is relatively difficult because some unrealistic details of the fake images are eliminated. Then, we randomly pick 202,599 fake images from the fake image pool. Finally, we have 400,198 training images and 5,000 test images containing real and fake images.

In the parameters setting of $D_1$ and $D_2$, the learning rate is 1e-3 and the maximum epoch is 15. The marginal value $m$ in contrastive loss is 0.5. Adam optimizer [11] is used for training $D_1$ and $D_2$. In first two epochs, we only adopt pairwise information to learn $D_1$. Afterward, the collected fake and real images will be used to learn $D_2$. The batch size is 32.

A traditional image forgery detection is used for comparison [3]. We also design a baseline approach based on the proposed

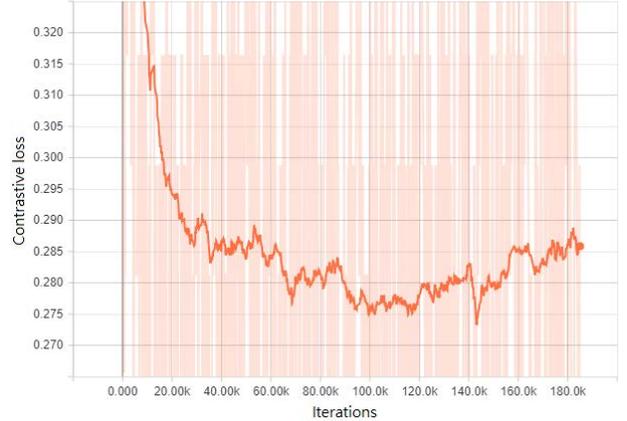

Fig. 2 The curve of the contrastive loss for learning $D_1$ using pairwise information.

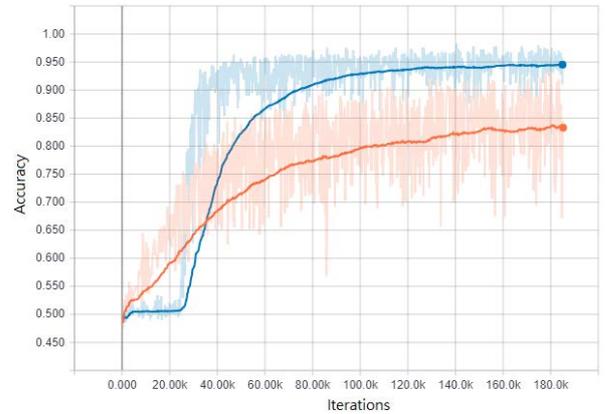

Fig. 3 The performance comparison between the proposed DeepFD with (Blue line) / without (Orang line) contrastive loss for training set excluding LSGAN.

DeepFD without contrastive loss to emphasize our contribution. On the other hand, we also remove the proposed classifier and replace it with other advanced classifiers such as RFC (Random forest classifier), SVM (Support vector machine), and LC (Linear classifier) to demonstrate the effectiveness of the proposed DeepFD.

#### B. Performance Comparison

To obtain the jointly discriminative feature for our task, we adopt pairwise information to guide D1 learning in two epochs. Figure 2 shows that the proposed contrastive loss is stable and will be converged. Although the contrastive loss is no longer used after 25,000 iterations (2 epochs), the contrastive loss did not increase. It has also verified a fact that the jointly discriminative feature learning is associated with classifier learning. Figure 3 presents the test accuracy comparison between the proposed DeepFD with and without contrastive loss learning. It is clear that the proposed DeepFD is easily converged and reach higher performance.

To demonstrate the effectiveness of the proposed DeepFD, we separate the fake images generated by one of the collected GAN methods from the training pool. A generalized DeepFD should be able to detect the fake images even they are not used in the training. Table II indicates the performance comparison between the proposed DeepFD and other baseline methods in terms of accuracy, precision, and recall. The proposed DeepFD significantly outperforms other techniques. It is also verified that the proposed DeepFD is more generalized and effective than others. On the other hand, we observed that the performance gain of the proposed method for the training dataset excluding LSGAN is significantly better because that the LSGAN shows fewer unrealistic details in the generated images. Therefore, the purely supervised approach (i.e., the proposed method without contrastive loss) cannot well capture the common features for the fake image. The proposed DeepFD is easier to extract the jointly discriminative feature for all kinds of the fake images, leading to higher performance.

*C. Visualization of the Unrealistic Details in the Fake Images*

Since the proposed DeepFD is designed to be a fully convolutional network, the feature maps can be visualized to localize the unrealistic details in the fake images. Inspired by fully convolutional network [14], there are two channels in the last convolutional layer in the proposed classifier $D_2$, leading to the learning goal of this layer will tend to learn a high activated feature representation, which can be considered as the unrealistic details localization. Six visualization results of the fake images generated by PGGAN are shown in Fig. 4. It well knows that the performance on image synthesis task of PGGAN is the state-of-the-art. However, some artifacts in the generated images still can be found. In the proposed DeepFD, most of such artifacts can be successfully localized, bring an extra benefit that the detecting performance can be further improved by investigating these visualized results.

## IV. CONCLUSIONS

In this study, we have proposed a novel deep forgery discriminator (DeepFD) based on embedding the contrastive loss, to successfully detect the fake / generated images generated by state-of-the-art GANs. To the best of our knowledge, the proposed approach is the first work to solve the problems of detecting the fake images. The main contribution is that the contrastive loss can be used to well capture the joint discriminative features of the fake images generated by different GANs. Besides, the proposed classifier refinement process leads to further boost the classification performance and can be used to visualize the unrealistic details in the fake images. Our experimental results demonstrate that the proposed method outperforms other baseline approaches in terms of precision and recall rate.

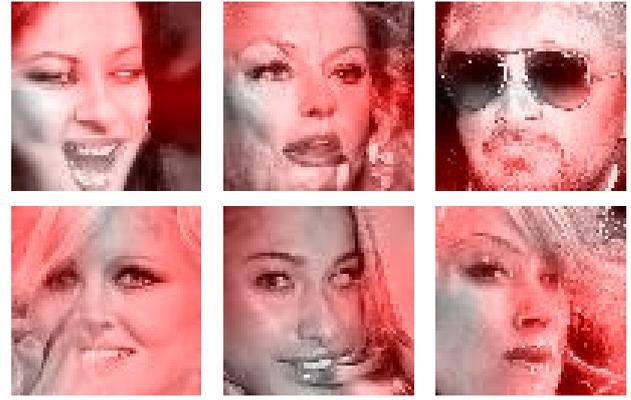

Fig. 4 The visualized results of the fake images generated by PGGAN. The region marked in red indicating the unrealistic regions that the proposed DeepFD detected.

ACKNOWLEDGMENT

This work was supported in part by the Ministry of Science and Technology of Taiwan under grant MOST 107-2218-E-020 -002 -MY3.

REFERENCES

[1] AI can now create fake porn, making revenge porn even more complicated, http://theconversation.com/ai-can-now-create-fake-porn-making-revenge-porn-even-more-complicated-92267, The conversation media group, 2018.
[2] C.C.Hsu, et al., "Video forgery detection using correlation of noise residue." in *Proc. of IEEE 10th Workshop on Multimedia Signal Processing*, 2008.
[3] H. Farid, "Image forgery detection," *IEEE Signal processing magazine* 26.2, pp. 16-25, 2009.
[4] Karras, Tero, et al. "Progressive growing of GANS for improved quality, stability, and variation," *arXiv preprint arXiv:1710.10196*, 2017.
[5] I. Goodfellow, et al. "Generative adversarial nets," in *Proc. of Advances in neural information processing systems*. 2014.
[6] A. Radford, et al.. "Unsupervised representation learning with deep convolutional generative adversarial networks," *arXiv preprint arXiv:1511.06434*, 2015.
[7] M. Arjovsky, et al., "Wasserstein gan," *arXiv preprint arXiv:1701.07875* (2017).
[8] Gulrajani, Ishaan, et al. "Improved training of wasserstein gans," *Advances in Neural Information Processing Systems*. 2017.
[9] X. Mao, et al. "Least squares generative adversarial networks," *2017 IEEE International Conference on Computer Vision (ICCV)*. IEEE, 2017.
[10] Y. LeCun, et al., "Handwritten digit recognition with a back-propagation network," *Advances in neural information processing systems*. 1990.
[11] I. Sutskever, et al. "On the importance of initialization and momentum in deep learning." *International conference on machine learning*. 2013.
[12] E. Simo-Serra, et al. "Discriminative learning of deep convolutional feature point descriptors," *Computer Vision (ICCV), 2015 IEEE International Conference on*. IEEE, 2015.
[13] Y.L. Chen and C.T. Hsu. "Detecting doubly compressed images based on quantization noise model and image restoration," in *Proc. of IEEE International Workshop on*. Multimedia Signal Processing, Oct. 2009.
[14] J. Long, et al. "Fully convolutional networks for semantic segmentation," in *Proc. of the IEEE conference on computer vision and pattern recognition*, pp.640-651, May 2016